\documentclass[twocolumn,pra,showpacs,floatfix]{revtex4}
\usepackage{dcolumn}                 % Aligning table columns
\newcolumntype{w}[1]{D{.}{.}{#1}}
\newcommand*{\centt}[1]{\multicolumn{1}{c}{#1}}
\newcommand*{\cent}[1]{\multicolumn{1}{c}{$#1$}}
\usepackage{times}
\usepackage{nicefrac}
\usepackage{amsmath}
\usepackage{amsfonts}
\usepackage{amssymb}
\usepackage{amsthm}

\begin{document}
\preprint{Version 2.0}

\title{Fine and hyperfine splitting of the $2P$ state in Li and Be$^+$}

\author{Mariusz Puchalski}
\email[]{mpuchals@fuw.edu.pl}

\author{Krzysztof Pachucki}
\email[]{krp@fuw.edu.pl}

\affiliation{Institute of Theoretical Physics, University of Warsaw,
             Ho\.{z}a 69, 00-681 Warsaw, Poland}

\begin{abstract}
Accurate calculations of the fine and hyperfine splitting
of the $2P$ state in Li and Be$^+$ isotopes using 
the explicitly correlated Hylleraas basis set are presented.
Theoretical predictions including the mixing of $P_{1/2}$ and $P_{3/2}$ states,
relativistic and quantum electrodynamics effects on hyperfine interactions,
are compared with experimental values. It is concluded that precise
spectroscopic determination of the nuclear magnetic moments
requires elimination of nuclear structure effects by combining measurements
for two different states.  
  
\end{abstract}

\maketitle

\section{Introduction}
The calculation of relativistic effects in the atomic structure is most often
performed with the explicit use of the Dirac equation, as in
relativistic configuration interaction \cite{yer_hfs}, 
many body perturbation theory \cite{der1_hfs}, 
relativistic coupled-cluster \cite{der2_hfs}, or multi-configuration 
Dirac-Fock \cite{bieron} methods. For light atomic systems
the more accurate approach is based on the expansion of the energy
in the fine structure constant $\alpha$. This method allows for 
a systematical inclusion of relativistic and quantum electrodynamics
(QED) contributions, as each correction can be expressed
in terms of the expectation value of some operator with the 
nonrelativistic wave function. 
With the use of explicitly  correlated basis functions, 
the nonrelativistic Schr\"odinger equations
for few electron systems can be solved very accurately.
The high precision is achieved also for relativistic and QED
corrections, provided more complicated integrals 
with inverse powers of inter-electronic distances can be performed.
Such calculations, which rely on expansion in $\alpha$,
have been performed for hydrogen and hydrogen-like ions 
up to the very high order of $m\,\alpha^8$ \cite{hydrogen}. 
Slightly lower precision was achieved for the helium fine structure
and for other helium energy levels, all terms up to
$m\,\alpha^6$ order have been obtained with approximate inclusion of dominant 
$m\,\alpha^7$ corrections \cite{h6hel}. For 3- and 4-electron atoms 
calculations have reached the order $m\,\alpha^5$ with 
partial inclusion of $m\,\alpha^6$ terms, which come 
from the electron self-energy.
The complete calculation of the $m\,\alpha^6$ contribution
for 3-electron systems has not been performed so far.

In this work we present accurate calculation of the fine and hyperfine
splittings in Li and Be$^+$ ions through $m\,\alpha^4$ and $m\,\alpha^5$
orders including the finite nuclear mass corrections.
Lithium fine structure have already been calculated in Hylleraas functions
by Yan and Drake in \cite{drake_fine}, but in a relatively small basis
and with the neglect of $P_{1/2}$ and $P_{3/2}$ mixing, which 
we find to play a significant role in the isotope shift. 
The hyperfine splitting
of $P$-states was calculated in many works using explicitly
relativistic methods \cite{yer_hfs, der1_hfs,der2_hfs, bieron} and
with the nonrelativistic multi-configuration Hartree-Fock 
method in \cite{yam,godefroid}.
We find by a comparison with our results,
that the most accurate previous calculation was that
performed by Yerokhin in \cite{yer_hfs}. For the 
comparison with experimental values we include $O(\alpha^2)$ relativistic corrections
from \cite{yer_hfs}, known $O(\alpha^2)$ QED
corrections and draw a conclusion that a largest uncertainty comes
from the not well known nuclear structure effects.

\section{Fine and hyperfine operators}
Let us briefly start with the description of the fine 
and hyperfine splitting in an arbitrary few electron atom.
The fine structure, neglecting relativistic $O(\alpha^2)$ corrections,
can be expressed as the expectation value
with the nonrelativistic wave function of the following operator
\begin{eqnarray}
H_{\rm fs} &=& \sum_a \frac{Z\,\alpha}{2\,r_a^3}\,\vec s_a\,
\biggl[\frac{(g-1)}{m^2}\,\vec r_a\times\vec p_a 
- \frac{g}{m\,m_{\rm N}}\,\vec r_a\times\vec p_{\rm N}\biggr]
\nonumber \\ &&+
\sum_{a\neq b}\frac{\alpha}{2\,m^2\,r_{ab}^3}\,
\vec s_a\bigl[g\,\vec r_{ab}\times\vec p_b
-(g-1)\,\vec r_{ab}\times\vec p_a\bigr]\nonumber \\
\end{eqnarray}
where $g$ is the free electron $g$-factor, which includes here
all QED corrections, $Z$ is the nuclear charge in units of 
the elementary charge $e$, $m$, $m_{\rm N}$ are the electron 
and nuclear masses respectively, finally $\vec s_a$ is 
the electron spin operator. For convenience of further calculations
we express $H_{\rm fs}$ in terms of $F_a^i$ and 
4 elementary operators $f_{a}^i$ in atomic units, namely  
\begin{eqnarray}
H_{\rm fs} &=&-i\,\sum_a \vec s_a\cdot\vec F_a,\nonumber \\
F_a^i &=& \varepsilon\,\biggl[\frac{Z\,(g-1)}{2}\,f_{1a}^i 
             + \frac{Z\,g}{2}\,\frac{m}{m_{\rm N}}\,f_{2a}^i
             + \frac{g}{2}\,f_{3a}^i\nonumber \\ &&
             - \frac{(g-1)}{2}\,f_{4a}^i\biggr],
\end{eqnarray}
where $\varepsilon =m\,\alpha^4$ and 
\begin{eqnarray}
\vec f_{1a} &=& \frac{\vec r_a}{r_a^3}\,\times\vec\nabla_a,\\
\vec f_{2a} &=& \frac{\vec r_a}{r_a^3}\times \sum_b \vec\nabla_b,\\
\vec f_{3a} &=& \sum_{b\neq a} \frac{\vec r_{ab}}{r^3_{ab}}\times\vec\nabla_b,\\
\vec f_{4a} &=& \sum_{b\neq a} \frac{\vec r_{ab}}{r^3_{ab}}\times\vec\nabla_a.
\end{eqnarray}

The hyperfine structure, neglecting relativistic $O(\alpha^2)$ corrections, 
is given by $H_{\rm hfs}$ operator in Eq. (\ref{08}).
We will treat the nucleus as any other particle
with mass $m_{\rm N}$ and with the $g$-factor $g_{\rm N}$ which is related to
the magnetic moment $\mu$ by the formula
\begin{equation}
g_{\rm N} = \frac{m_{\rm N}}{Z\,m_{\rm p}}\,\frac{\mu}{\mu_{\rm N}}\,\frac{1}{I},
\end{equation}
where $\mu_{\rm N}$ is the nuclear magneton and $I$ is the nuclear spin.
Nuclear masses, spins, magnetic dipole and electric quadrupole moments
of Li and Be isotopes are taken from literature and 
are all presented in Table \ref{table1}.
\squeezetable
\begin{table*}[!hbt]
\renewcommand{\arraystretch}{1.0}
\caption{Data for Lithium and Beryllium isotopes.
 Atomic binding energy of $ E_{\rm Li} =-7.281$ au, $E_{\rm Be}=-14.669$ au.
The value for the quadrupole moment of $^7$Be is a theoretical estimate \cite{7be_quad}.}
\label{table1}
\begin{ruledtabular}
\begin{tabular}{lw{2.14}llw{2.12}lw{2.6}lw{2.6}l}
& \centt{atomic mass [u]} & \centt{Ref.} & \cent{I^\pi}  & \cent{\mu[\mu_N]}
& \centt{Ref.} &\cent{Q[{\rm fm}^2]} & \centt{Ref.} & \cent{r_E} & \centt{Ref.}\\
\hline
$^6$Li     &6.015122794(16)  &\cite{li_mass}&$1^+$  &0.822\,047\,3(6)&\cite{raghavan,stone}&-0.0806(6)&\cite{ceder}&2.540(28)&\cite{lit_iso2}\\
$^7$Li     &7.0160034256(45) &\cite{7limass}&$3/2^-$&3.256\,426\,8(17)& \cite{raghavan,stone}&-4.00(3)&\cite{voelk}&2.390(30)&\cite{jager}\\
$^8$Li     &8.02248624(12)      &\cite{li_mass}&$2^+$  &1.653560(18)& \cite{raghavan,stone} &+3.14(2)&\cite{borr}&2.281(32)&\cite{lit_iso2}\\
$^9$Li     &9.02679020(21)      &\cite{li_mass}&$3/2^-$&3.43678(6)&\cite{borr}&-3.06(2)&\cite{borr}&2.185(33)&\cite{lit_iso2}\\
$^{11}$Li  &11.04372361(69)     &\cite{li_mass}&$3/2^-$&3.6712(3)&\cite{neugart}&-3.33(5)&\cite{neugart}&2.426(34)&\cite{lit_iso2}\\
$^7$Be     &7.016\,929\,83(11)  &\cite{nu_mass}&$3/2^-$&-1.39928(2)&\cite{okada}&-6.11&\cite{arai}&2.646(14)&\cite{be_iso} \\
$^9$Be     & 9.012\,182\,20(43) &\cite{nu_mass}&$3/2^-$&-1.177\,432(3)&\cite{win,itano}&-5.288(38)&\cite{sundholm}&2.519(12)&\cite{jansen} \\
$^{10}$Be   &10.013\,533\,82(43)&\cite{nu_mass}&$0^+$&&&&&2.358(16)&\cite{be_iso} \\
$^{11}$Be   &11.021\,661\,55(63)&\cite{11bemass}&$1/2^+$&-1.681\,3(5)& \cite{kapp,be_iso}&&&2.463(16)&\cite{be_iso} \\
$^{12}$Be   &12.026\,921(16)    &\cite{nu_mass}&$0^+$& \\
$^{14}$Be   &14.042\,890(140)   &\cite{nu_mass}&$0^+$& \\
\end{tabular}
\end{ruledtabular}
\end{table*}
With the help of $g_{\rm N}$ $H_{\rm hfs}$ can be written as
\begin{eqnarray}
H_{\rm hfs} &=&
\sum_a\biggl[
\frac{2}{3}\,\frac{Z\,\alpha\,g\,g_{\rm N}}{m\,m_{\rm N}}\,
\vec s_a\cdot\vec I\,\pi\,\delta^3(r_a)
\nonumber \\ &&
-\frac{Z\,\alpha\,g\,g_N}{4\,m\,m_N}\,
\frac{s_a^i\,I^j}{r_a^3}\,
\biggl(\delta^{ij}-3\frac{r_a^i\,r_a^j}{r_a^2}\biggr)
\nonumber \\ &&\hspace{-1ex}
+\frac{Z\,\alpha\,g_N}
{2\,m\,m_N}\,\vec I\cdot\frac{\vec r_a}{r_a^3}\times\vec p_a
-\frac{Z\,\alpha\,(g_N-1)}
{2\,m_N^2}\,\vec I\cdot\frac{\vec r_a}{r_a^3}\times\vec p_N
\nonumber \\ &&
+\frac{Q}{6}\,\frac{\alpha}{r_a^3}\,
\biggl(\delta^{ij}-3\frac{r_a^i\,r_a^j}{r_a^2}\biggr)\,
\frac{3\,I^i\,I^j}{I\,(2\,I-1)}\biggr] \label{08}\\
&\equiv&
\vec I\cdot\vec G + \frac{H^{ij}}{6}\,\frac{3\,I^i\,I^j}{I\,(2\,I-1)},
\end{eqnarray}
where $Q$ is the electric quadrupole moment. 
For convenience of further calculations we express $H_{\rm hfs}$
in terms of $H_a$, $H_a^{ij}$, $H^i$, and $H^{ij}$ operators, namely
\begin{eqnarray}
G^i &=&\sum_a s_a^i\,H_a + \sum_a s_a^j\,H_a^{ij}-i\,H^i,\\
H_a &=& \varepsilon\,Z\,g_N\,\frac{m}{m_N}\,\frac{g}{6}\,h_a, \\
H_a^{ij} &=& -\varepsilon\,Z\,g_N\,\frac{m}{m_N}\,\frac{g}{4}\,h_a^{ij},\\
H^i &=& \varepsilon\,\biggl[\frac{Z}{2}\,g_N\,\frac{m}{m_N}\,h_1^i - 
        \frac{Z}{2}\,(g_N-1)\,\frac{m^2}{m_N^2}\,h_2^i\biggr],\\
H^{ij} &=& \varepsilon\,m^2\,Q\,h^{ij},
\end{eqnarray} 
where $h$ operators (in atomic units) are
\begin{eqnarray}
\vec h_1    &=& \sum_a \frac{\vec r_a}{r_a^3}\,\times\vec\nabla_a,\\
\vec h_2    &=& \sum_a \frac{\vec r_a}{r_a^3}\,\times\sum_b\vec\nabla_b,\\
h_a         &=& 4\,\pi\,\delta^3(r_a),\\
h_a^{ij}    &=& \frac{1}{\,r_a^3}\,
                \biggl(\delta^{ij}-3\,\frac{r_a^i\,r_a^j}{r_a^2}\biggr),\\
h^{ij}       &=& \sum_a \frac{1}{r_a^3}\,
            \biggl(\delta^{ij}-3\,\frac{r_a^i\,r_a^j}{r_a^2}\biggr).
\end{eqnarray}

\section{Matrix elements}
Matrix elements of the fine and hyperfine operators are evaluated
with the nonrelativistic wave function. This function is obtained by solving
the Schr\"odinger equation in the 3-electron Hylleraas
basis set. Finite nuclear mass corrections are included by reduced mass
scaling and perturbative treatment of the mass polarization correction. 
All matrix elements are expressed in terms of Hylleraas integrals, which
are obtained with the help of recursion relations \cite{rec1, rec2, li_ground, rec3}. 
The high accuracy is achieved by the use of a large number 
of about 15000 Hylleraas functions, and we have already demonstrated
the advantages of this approach by the calculation of the isotope shift
in Li \cite{mp_li_iso} and Be$^+$ ions \cite{lit_rel}.

The nonrelativistic wave function is the antisymmetrized product 
of spacial and spin functions of the form
\begin{eqnarray}
\psi^i_{a} &=& {\cal A}[\phi_a^i(\vec r_1,\vec r_2,\vec r_3)\,\chi]\,,\\
\phi^i_a(\vec r_1, \vec r_2, \vec r_3) &=& r_a^i\,
e^{-w_1\,r_1-w_2\,r_2-w_3\,r_3}\nonumber \\ &&
r_{23}^{n_1}\,r_{31}^{n_2}\,r_{12}^{n_3}\,r_{1}^{n_4}\,r_{2}^{n_5}\,r_{3}^{n_6}\,,\\
\chi &=& [\alpha(1)\,\beta(2)-\beta(1)\,\alpha(2)]\,\alpha(3),
\end{eqnarray}
where $\sigma_z\,\alpha(.) = \alpha(.) $ and $\sigma_z\,\beta(.) = -\beta(.) $.
Matrix elements of each operator, after eliminating spin variables can take
the standard form
\begin{eqnarray}
\langle i|H|j\rangle_S &\equiv& \bigl\langle \phi'{}^i(r_1,\,r_2,\,r_3)|H|
\nonumber \\ &&
2\,\phi^j(r_1,r_2,r_3)+2\,\phi^j(r_2,r_1,r_3)
\nonumber \\ &&
-\phi^j(r_2,r_3,r_1)-\phi^j(r_3,r_2,r_1)\nonumber \\ &&
-\phi^j(r_3,r_1,r_2)-\phi^j(r_1,r_3,r_2)
\bigr\rangle
\end{eqnarray}
and, what we call, the Fermi form
\begin{eqnarray}
\langle i|H_a|j\rangle_F &\equiv& \langle \phi'{}^i(r_1,\,r_2,\,r_3)|
2\,H_3\,[\phi^j(r_1,r_2,r_3)
\nonumber \\ &&
+\phi^j(r_2,r_1,r_3)]-(H_1-H_2+H_3)
\nonumber \\ &&
\times[\phi^j(r_2,r_3,r_1)+\phi^j(r_3,r_2,r_1)]
\nonumber \\ &&
-(H_2-H_1+H_3)\,[\phi^j(r_1,r_3,r_2)
\nonumber \\ &&
+\phi^j(r_3,r_2,r_1)]\rangle,
\end{eqnarray}
with the assumption that the norm is $\sum_{i=1}^3\langle i|i\rangle_S =1$.
The matrix element of the fine structure Hamiltonian becomes
\begin{eqnarray}
\langle H_{\rm fs} \rangle_J &=&
\langle -i\,\sum_a\vec s_a\cdot\vec F_a\rangle_J \nonumber \\ &=&
\epsilon^{ijk}\,\langle i|F_a^j|k\rangle_F\,
\left\{
\begin{array}{r}
\frac{1}{2},\; J=1/2\\
- \frac{1}{4},\; J=3/2
\end{array}\right.
\end{eqnarray}
and the fine splitting is
\begin{equation}
E_{\rm fs} = \langle H_{\rm fs} \rangle_{3/2} - \langle H_{\rm fs} \rangle_{1/2}
          = -\frac{3}{4}\,\epsilon^{ijk}\,\langle i|F_a^j|k\rangle_F.
\end{equation}
The matrix elements of the hyperfine structure Hamiltonian takes the form
\begin{eqnarray}
\langle H_{\rm hfs} \rangle_J &=&
\biggl\langle  \vec I\cdot\vec G + 
\frac{3\,I^i\,I^j}{I\,(2\,I-1)}\,\frac{H^{ij}}{6}\biggr\rangle
 \\ &=&
A_J\,\vec I\cdot\vec J + \frac{B_J}{6}\,
\frac{3\,(I^i\,I^j)^{(2)}}{I\,(2\,I-1)}\,
\frac{3\,(J^i\,J^j)^{(2)}}{J\,(2\,J-1)},\nonumber
\end{eqnarray}
where $A_J$ and $B_J$ are magnetic dipole and electric quadrupole
hyperfine constants. They are all expressed in terms of standard and Fermi
matrix elements, namely 
\begin{eqnarray}
A_J &=& \frac{1}{J\,(J+1)}\,\langle\vec J\cdot\vec G\rangle_J, \\
A_{1/2} &=& 
        -\frac{1}{3}\,\langle k|H_a|k\rangle_F
        -\frac{2}{3}\,\epsilon^{ijk}\,\langle i|H^j|k\rangle_S
        +\frac{2}{3}\,\langle i|H_a^{ij}|j\rangle_F \nonumber \\
A_{3/2} &=& \frac{1}{3}\,\langle k|H_a|k\rangle_F
        -\frac{1}{3}\,\epsilon^{ijk}\,\langle i|H^j|k\rangle_S
        -\frac{1}{15}\,\langle i|H_a^{ij}|j\rangle_F\nonumber\\[1ex]     
B_J &=& \frac{2}{(2\,J+3)\,(J+1)}\,\langle J^i J^j\,H^{ij}\rangle_J,\\ 
B_{1/2} &=& 0, \nonumber\\
B_{3/2} &=&-\frac{1}{5}\,\langle i|H^{ij}|j\rangle_S.\nonumber
\end{eqnarray}
Numerical values for all matrix elements involved in these calculations are
presented in Table \ref{table2}.  They have been obtained by extrapolation
to infinite basis set and uncertainties reflect the numerical convergence.
\begin{table*}[!hbt]
\renewcommand{\arraystretch}{1.0}
\caption{Matrix elements in atomic units of operators involved 
         in the fine and hyperfine splitting
         of P-states, infinite mass and the mass polarization correction with the 
         coefficient $-m/(m+m_{\rm N})$. $\langle k|h_{a}|k\rangle_F$
         corresponds to $a_c$ from Ref. \cite{yer_hfs},
         $\langle i|h_{a}^{ij}|j\rangle_F$ to $10\,a_{sd}$,
         $\epsilon^{ijk}\,\langle i|h_{1}^j|k\rangle_S$ to $2\,a_l$,
         and $\langle i|h^{ij}|j\rangle_S$ to $b_q/2$. Numerical uncertainties
         are due to extrapolation to the infinite basis set and reflect the
         numerical convergence.}
\label{table2}
\begin{ruledtabular}
\begin{tabular}{rw{2.12}w{2.12}w{2.12}w{2.12}l}
operator & \centt{Li($2P$)$_\infty$} & \centt{mass pol. corr.}  
         & \centt{Be$^+(2P)_\infty$} & \centt{mass pol. corr.} & \centt{Ref.}\\
\hline
$\epsilon^{ijk}\,\langle i|f_{1a}^j|k\rangle_F$  
 &-0.125\,946\,353\,2(18) & 0.376\,388(3) &-0.969\,131\,4 (8) &  3.043\,395 (9) & \\
$\epsilon^{ijk}\,\langle i|f_{2a}^j|k\rangle_F$  
 & 0.022\,524\,89(15) &            & 0.339\,008\,2(2) &   &\\
$\epsilon^{ijk}\,\langle i|f_{3a}^j|k\rangle_F$  
 & 0.038\,473\,58(12) &-0.213\,52(3) & 0.360\,851\,6(2) & -1.549\,82(12)  &\\
$\epsilon^{ijk}\,\langle i|f_{4a}^j|k\rangle_F$  
 &-0.224\,640\,68(6) & 0.570\,582(6) &-1.659\,492\,5(2) &  4.532\,62(13)  &\\
$\langle k|h_{a}|k\rangle_F$                    
 &-0.214\,620\,4(19) & 2.376\,4(5) &-1.083\,916\,1(8) &12.232(12)  &\\
 &-0.214\,67           &             &-1.084\,2 & &\cite{yer_hfs} \\
 &-0.214\,78(5)        &             &        & &\cite{drake_hfs} \\
$\langle i|h_{a}^{ij}|j\rangle_F$               
 &-0.134\,775\,3(5) & 0.357\,1(17) &-1.026\,978 (3) & 2.775\,0 (8)  &\\
 &-0.134\,77        &             &-1.026\,9 & &\cite{yer_hfs} \\
$\epsilon^{ijk}\,\langle i|h_{1}^j|k\rangle_S$  
 &-0.126\,256\,153(17) & 0.400\,67(5) &-0.970\,443\,9 (3) & 3.116\,48 (5) & \\
 &-0.126\,250         &             &-0.970\,32 & &\cite{yer_hfs} \\
$\epsilon^{ijk}\,\langle i|h_{2}^j|k\rangle_S$   
 &0.044\,419\,16 (19)&             & 0.398\,663\,5 (7) &   &\\
$\langle i|h^{ij}|j\rangle_S$                   
 &-0.113\,097(2) & 0.334\,9(9) &-0.918\,134(3) & 2.628(3)  &\\
 &-0.113\,085    &             &-0.918\,10 & &\cite{yer_hfs} 
\end{tabular}
\end{ruledtabular}
\end{table*}
Matrix elements of the fine structure operators have been derived previously by Yan and Drake in
\cite{drake_fine} and later by us in \cite{lit_rel}. Small differences with
results of \cite{drake_fine} come from the not very large number of basis functions
used in that work. The hyperfine operators have been
previously obtained in several works, i.e. \cite{yer_hfs, godefroid, fcpc}
and we compare our result with the most accurate one from \cite{yer_hfs}
with which we agree well. 

\section{The second order contribution}
The hyperfine Hamiltonian $H_{\rm hfs}$ mixes $2^2P_{1/2}$ with  $2^2P_{3/2}$
what leads to additional contributions to fine and hyperfine 
splittings \cite{derevianko}.
Since this mixing is not very large one can use the second order 
perturbative formula which involves off-diagonal matrix elements
\begin{eqnarray}
&&\delta E(P_{1/2})_{m_1m_2} = \label{30}\\&&  
\sum_m\frac{\langle P_{1/2},m_1| H_{\rm hfs}|P_{3/2},m\rangle\,
\langle P_{3/2},m| H_{\rm hfs}|P_{1/2},m_2\rangle}{E(P_{1/2})-E(P_{3/2})} \nonumber\\
&&\delta E(P_{3/2})_{m_1m_2} =  \nonumber\\&& 
\sum_m\frac{\langle P_{3/2},m_1| H_{\rm hfs}|P_{1/2},m\rangle\,
\langle P_{1/2},m| H_{\rm hfs}|P_{3/2},m_2\rangle}{E(P_{3/2})-E(P_{1/2})}\nonumber
\end{eqnarray}
To calculate them one can use Clebsch-Jordan coefficients 
and Racah algebra \cite{johnson}. In the simpler approach presented 
here, we introduce the operator $K$, such that
$\langle J,m|\vec K|J,m'\rangle = 0$ for $J=1/2, 3/2$, 
but does not change $L$ nor $S$, namely
\begin{eqnarray}
\vec K &=& \vec S-\vec J\,\biggl(\frac{1}{2}-\frac{5}{8\,J\,(J+1)}\biggr)
\nonumber \\
&=&\left\{
\begin{array}{r}
\vec S+\frac{1}{3}\,\vec J,\;J=1/2\\
\vec S-\frac{1}{3}\,\vec J,\;J=3/2
\end{array}\right.
\end{eqnarray}
Then the off-diagonal matrix elements can be transformed to the form
\begin{align}
&\langle P_J,m|H_{\rm hfs}|P_{J'},m'\rangle \nonumber \\
&=I^i\,\langle P_J,m|G^i|P_{J'},m'\rangle 
+\frac{3\,I^i\,I^j}{I\,(2\,I-1)}\,
\frac{\langle P_J,m|H^{ij}|P_{J'},m'\rangle}{6}
\nonumber \\ 
&= I^i\,X\,\langle J,m|K^i|J',m'\rangle +
\frac{3\,I^i\,I^j}{I\,(2\,I-1)}\,\frac{Y}{6}
\nonumber \\&\;\;\;\;\times
\langle J,m|(L^i\,L^j)^{(2)}|J',m'\rangle
\end{align}
with $X$ and $Y$ coefficients being
\begin{eqnarray}
X &=& \langle k|H_a|k\rangle_F + 
      \frac{\epsilon^{ijk}}{2}\,\langle i|H^j|k\rangle_S +
      \frac{1}{4} \langle i|H_a^{ij}|j\rangle_F \\
Y &=&-\frac{3}{5}\,\langle i|H^{ij}|j\rangle_S.
\end{eqnarray}
The second order correction to energy 
due to $H_{\rm hfs}$ in Eqs. (\ref{30}) neglecting the small $Y^2$ term becomes
\begin{eqnarray}
\delta E(P_{1/2}) &=& -\frac{X^2}{E_{\rm fs}}\,I^i\,I^j\,
\langle K^i\,K^j\rangle_{J=1/2} 
-\frac{X\,Y}{E_{\rm fs}}
\nonumber \\ && \times\frac{I^k\,I^i\,I^j}{I\,(2\,I-1)}\,
\langle K^k\,(L^i\,L^j)^{(2)}\rangle_{J=1/2}\nonumber \\
&=& -\frac{X^2}{E_{\rm fs}}\,\frac{2}{9}\,
    \bigl(\vec I^{\,2}+\vec I\cdot\vec J\bigr)
+\frac{X\,Y}{E_{\rm fs}}\,\frac{2\,I+3}{9\,I}\,\vec I\cdot\vec J,
\nonumber \\ \\
\delta E(P_{3/2}) &=& \frac{X^2}{E_{\rm fs}}\,\,I^i\,I^j\,
\langle K^i\,K^j\rangle_{J=3/2} 
+\frac{X\,Y}{E_{\rm fs}}\,\frac{I^k\,I^i\,I^j}{I\,(2\,I-1)}
\nonumber \\ && \times
\langle K^k\,(L^i\,L^j)^{(2)}\rangle_{J=3/2}
\nonumber \\
&=& \frac{X^2}{E_{\rm fs}}\,\frac{1}{9}\,
\bigl[\vec I^{\,2}-\vec I\cdot\vec J-(I^i\,I^j)^{(2)}
\,(J^i\,J^j)^{(2)}\bigr]\nonumber \\ &&
+\frac{X\,Y}{E_{\rm fs}}\,
\biggl[-\frac{(2\,I+3)}{90\,I}\,\vec I\cdot\vec J
+\frac{1}{18}\,\frac{3\,(I^i\,I^j)^{(2)}}{I\,(2\,I-1)}
\nonumber \\ && \times (J^i\,J^j)^{(2)}\biggr],
\end{eqnarray}
where we omitted the magnetic octupole coupling, 
the so called $C_J$ coefficient.
Resulting corrections to the fine and hyperfine splittings are
\begin{eqnarray}
\delta E_{\rm fs} &=&\frac{X^2}{E_{\rm fs}}\,\frac{I\,(I+1)}{3},\\
\delta A_{1/2} &=& -\frac{2}{9}\,\frac{X^2}{E_{\rm fs}}
+\frac{2\,I+3}{9\,I}\,\frac{X\,Y}{E_{\rm fs}},\\
\delta A_{3/2} &=& -\frac{1}{9}\,\frac{X^2}{E_{\rm fs}}
-\frac{2\,I+3}{90\,I}\,\frac{X\,Y}{E_{\rm fs}},\\
\delta B_{3/2} &=& -\frac{2\,I\,(2\,I-1)}{9}\,\frac{X^2}{E_{\rm fs}}
+\frac{1}{3}\,\frac{X\,Y}{E_{\rm fs}}.
\end{eqnarray}

\section{Results}
Numerical results for the fine splitting in Li and Be$^+$ isotopes
are shown in Table \ref{table3}. $E^{(0)}_{\rm fs}$ is the leading
contribution with the exact electron $g$-factor, but in the 
infinite nuclear mass limit, $E^{(1)}_{\rm fs}$ 
is the finite nuclear mass correction, and $\delta E_{\rm fs}$
is the $P_{1/2}-P_{3/2}$ mixing term. The higher order relativistic and 
QED corrections are not known, as they have not yet been evaluated.
Finally $\Delta E_{\rm fs}$ is the isotope shift
with respect to $^7$Li and $^9$Be$^+$. Our result for
this isotope shift in the fine structure $\Delta E_{\rm fs}$ of Li 
differs significantly from the previous calculations in \cite{drake_fine} 
due to the inclusion of the important second order contribution $\delta E_{\rm fs}$. 
However, it differs also from all the experimental values, see Table \ref{table3}.

\begin{table*}[!hbt]
\renewcommand{\arraystretch}{1.0}
\caption{Fine splitting of 2P-states in Li and Be$^+$ isotopes in MHz with
$\varepsilon = 2\,R\,c\,\alpha^2 = 6\;579\;683\;921 $ MHz.
$\Delta E_{\rm fs}$ is the isotope shift with respect to $^7$Li and $^9$Be.
It is not clear whether the experimental value of Orth {\em et al.} \cite{orth1} 
for the $^7$Li fine structure includes $\delta E_{\rm fs}$ due to their
diagonal and of-diagonal parametrization of hyperfine matrix elements.}
\label{table3}
\begin{ruledtabular}
\begin{tabular}{rw{6.7}w{6.7}w{6.7}w{6.7}w{6.7}l}
         & \centt{$^6$Li} & \centt{$^7$Li} & \centt{$^8$Li} & \centt{$^9$Li} & \centt{$^{11}$Li} &\centt{Ref.}\\
$E_{\rm fs}^{(0)}$            & 10\,053.707\,2(83) & 10\,053.707\,2(83) & 10\,053.707\,2(83) & 10\,053.707\,2(83) & 10\,053.707\,2(83) \\
$E_{\rm fs}^{(1)}$              &    -2.786\,8(6)  &      -2.389\,1(5)  &      -2.089\,3(4)  &    -1.856\,8(4)    &      -1.517\,7(3)  \\
$\delta E_{\rm fs}$            &     0.012\,17    &       0.159\,16    &       0.036\,93    &     0.177\,23      &       0.202\,21 \\
$\Delta E_{\rm fs}$            &    -0.544\,7(1)   & 10\,051.477(8)    & 0.177\,6(1)    & 0.550\,4(1) & 0.914\,5(2)\\[1ex]
                              &    -0.396         & 10\,051.235(12)   & 0.298    & 0.529 & 0.851 & \cite{drake_be, drake_fine}\\
expt.                         & 0.863(79) & 10\,053.184(58) &&&& \cite{das2,orth1}\\
expt.                         & -0.155(77)& 10\,053.39(21)  &&&& \cite{noble,scherf}\\[1ex]
        & \centt{$^7$Be$^+$}  & \centt{$^9$Be$^+$} &\centt{$^{10}$Be$^+$} &  \centt{$^{11}$Be$^+$} & \centt{$^{14}$Be$^+$} \\
$E_{\rm fs}^{(0)}$              & 197\,039.150(81)   & 197\,039.150(81)   & 197\,039.150(81) & 197\,039.150(81)  & 197\,039.150(81)    \\
$E_{\rm fs}^{(1)}$              &      -27.320(3)    &      -21.270(2)    &      -19.141(2)  &       -17.391(2)  &      -13.649\,2(15) \\
$\delta E_{\rm fs}$             &        0.045\,56   &        0.032\,25   &      0.000      &      0.118\,34  & 0.000            \\
$\Delta E_{\rm fs}$             & -6.037(1)   & 197017.727(21) & 2.097(1) & 3.965(1) & 7.589(1)\\[1ex]
                               & -6.049       &                & 2.13     &   3.878    & & \cite{drake_be}       
\end{tabular}
\end{ruledtabular}
\end{table*}

\begin{table*}[!hbt]
\renewcommand{\arraystretch}{1.0}
\caption{Hyperfine splitting of the $2P$-states in Li isotopes in MHz. Results of
  Yerokhin \cite{yer_hfs} are corrected by inclusion of $\delta A$ and $\delta
  B$, and by the use of more accurate electric quadrupole moments for $^6$Li
  and $^7$Li. Results of Orth {\em et al.} \cite{orth1,orth2} for $A$ and $B$
  constants in $^7$Li are shifted by $\delta A$ and $\delta B$,  as these authors 
  parametrized results of their measurement by diagonal and of-diagonal
  parts separately. Uncertainties of final theoretical predictions are due to
  higher order corrections and the approximate treatment of the nuclear structure contribution.
  Not shown are uncertainties due to inaccuracies of magnetic dipole and electric
  quadrupole moments.}
\label{table4}
\begin{ruledtabular}
\begin{tabular}{rw{6.6}w{6.6}w{6.6}w{6.6}w{6.8}l}
& \centt{$^6$Li} & \centt{$^7$Li} & \centt{$^8$Li} & \centt{$^9$Li} &\centt{$^{11}$Li} & \centt{Ref.} \\
$A_{1/2}^{\rm nrel}$  &    17.404\,70(4) &      45.963\,37(11)&      17.504\,24(4) &    48.507\,52(11)  &      51.815\,18(12)\\
$\delta A_{1/2}$      &    -0.004\,05    &      -0.027\,29    &      -0.004\,37    &    -0.030\,69      &      -0.035\,00    \\
                      &    -0.004\,01    &      -0.027\,0     &                    &                    &            &\cite{derevianko}\\
$A_{1/2}^{\rm rel}$   &     0.003\,53    &       0.009\,32    &       0.003\,55    &     0.009\,84      &       0.010\,51  &\cite{yer_hfs}  \\
$A_{1/2}^{\rm qed}$   &    -0.001\,08    &      -0.002\,86    &      -0.001\,09    &    -0.003\,01      &      -0.003\,22    \\
$A_{1/2}^{\rm fns}$   &    -0.001\,36    &      -0.003\,39    &      -0.001\,23    &    -0.003\,27   &      -0.003\,88   \\
$A_{1/2}$             &    17.401\,7(4)  &      45.939\,2(11)  &      17.501\,1(4)  &    48.480\,4(11)    &      51.783\,6(13)  \\[1ex]
                      &    17.401\,8(5)  &      45.939(1)     &&&&\cite{yer_hfs}\\
expt.                 &    17.371(18)    &      45.887 (25)   &&&&\cite{orth2, orth1}\\
expt.                 &    17.386(31)    &      46.010(25)    &&&&\cite{walls}\\
expt.                 &    17.394(4)     &      46.024(3)     &&&&\cite{das1}\\[2ex]
$A_{3/2}^{\rm nrel}$  &    -1.152\,35(2) &      -3.042\,14(4) &      -1.158\,31(2) &    -3.209\,24(4)   &      -3.427\,19(4) \\
$\delta A_{3/2}$      &    -0.002\,03    &      -0.014\,25    &      -0.002\,03    &    -0.015\,84      &      -0.018\,07    \\
                      &    -0.002\,01    &      -0.014\,1     &&&            &\cite{derevianko}\\
$A_{3/2}^{\rm rel}$   &    -0.001\,84    &      -0.004\,85    &      -0.001\,85    &    -0.005\,12      &      -0.005\,46 &\cite{yer_hfs}   \\
$A_{3/2}^{\rm qed}$   &     0.001\,08    &       0.002\,86    &       0.001\,09    &     0.003\,01      &       0.003\,22    \\
$A_{3/2}^{\rm fns}$   &     0.001\,36   &       0.003\,39    &       0.001\,23  &     0.003\,27   &       0.003\,88  \\
$A_{3/2}$             &    -1.153\,7(4)  &      -3.055\,0(11)  &      -1.159\,8(4)  &    -3.223\,8(11)    &      -3.443\,6(13) \\[1ex]
                      &    -1.155\,0(5)  &      -3.058(1)     &&&&\cite{yer_hfs}\\
expt.                 &    -1.157(8)     &      -3.069(14)    &&&&\cite{orth2,orth1}\\[2ex]
$B_{3/2}^{\rm nrel}$  &    -0.004\,28    &      -0.212\,59    &       0.166\,88    &    -0.162\,63      &      -0.176\,98    \\
$\delta B_{3/2}$      &    -0.004\,05    &      -0.084\,14    &      -0.024\,85    &    -0.093\,91      &      -0.107\,13    \\
                      &    -0.004\,02    &      -0.083\,4     &                    &               &            &\cite{derevianko}\\
$B_{3/2}^{\rm rel}$   &     0.000\,00    &       0.000\,02    &      -0.000\,01    &     0.000\,01      &       0.000\,01 &\cite{yer_hfs}  \\
$B_{3/2}$             &    -0.008\,33    &      -0.296\,71(8) & 0.142\,02(2) &    -0.256\,53(9)   &      -0.284\,10(11)\\[1ex]
                      &    -0.008\,33   &      -0.296\,69(2)    &&&&\cite{yer_hfs}\\
expt.                 &    -0.014(14)     &      -0.305(29)    &&&&\cite{orth2, orth1}
\end{tabular}
\end{ruledtabular}
\end{table*}

\begin{table}[!hbt]
\renewcommand{\arraystretch}{1.0}
\caption{Hyperfine splitting of 2P-states in Be$^+$ isotopes in MHz. Results of
  Yerokhin \cite{yer_hfs} are corrected by inclusion of $\delta A$ and $\delta
  B$. Uncertainties of final theoretical predictions are due to
  higher order corrections and the approximate treatment of the nuclear structure contribution.
  Not shown are uncertainties due to inaccuracies of magnetic dipole and electric
  quadrupole moments.}
\label{table5}
\begin{ruledtabular}
\begin{tabular}{rw{4.6}w{4.6}w{4.7}l}
& \centt{$^7$Be$^+$} & \centt{$^9$Be$^+$} & \centt{$^{11}$Be$^+$} &\centt{Ref.} \\
$A_{1/2}^{\rm nrel}$   &     -140.069\,6(3) &     -117.859\,2(3) &      -504.874\,5(8)  \\
$\delta A_{1/2}$       &       -0.009\,61   &       -0.006\,83   &        -0.105\,20    \\
$A_{1/2}^{\rm rel}$    &       -0.096\,8    &       -0.081\,5    &        -0.349 &\cite{yer_hfs} \\
$A_{1/2}^{\rm qed}$    &        0.008\,26   &        0.006\,95   &         0.029\,75    \\
$A_{1/2}^{\rm fns}$    &        0.010\,85   &        0.008\,69   &         0.055\,50    \\ 
$A_{1/2}$              &     -140.157(3)    &     -117.932(3)    &      -505.245(16)    \\[1ex]
                       &                    &     -117.926(4)    &&\cite{yer_hfs}\\
expt.                  &     -140.17(18)    &     -118.00(4)     &       -505.41(5)&\cite{be_iso}    \\
expt.                  &                    &      118.6(36)     &&\cite{boll} \\[2ex]
$A_{3/2}^{\rm nrel}$   &       -1.215\,33(2)&       -1.024\,81(2)&        -4.395\,48(8)  \\
$\delta A_{3/2}$       &       -0.003\,90   &       -0.002\,76   &        -0.052\,60     \\
$A_{3/2}^{\rm rel}$    &        0.020\,3    &        0.017\,1    &         0.073\,5 &\cite{yer_hfs}  \\
$A_{3/2}^{\rm qed}$    &       -0.008\,26   &       -0.006\,95   &        -0.029\,75     \\
$A_{3/2}^{\rm fns}$    &       -0.010\,85   &       -0.008\,69   &        -0.055\,50      \\
$A_{3/2}$              &       -1.218(3)    &       -1.026(3)    &        -4.460(16)     \\[1ex]
                       &                    &       -1.018(3)    &&\cite{yer_hfs}\\[2ex]
$B_{3/2}^{\rm nrel}$   &       -2.636\,19(1)&       -2.281\,54(1)&                       \\
$\delta B_{3/2}$       &       -0.025\,43   &       -0.018\,03   &                       \\
$B_{3/2}^{\rm rel}$    &        0.000\,20   &        0.000\,17   &&\cite{yer_hfs}  \\
$B_{3/2}$              &       -2.661\,42(3)&       -2.299\,40(3)&      \\[1ex]
                       &                    &      -2.299\,25(17) & & \cite{yer_hfs}
\end{tabular}
\end{ruledtabular}
\end{table}
Numerical values of all significant contributions to the hyperfine constants
of the $2^2P_{1/2}$ and $2^2P_{3/2}$ states in Li and Be$^+$ isotopes are shown in Tables
\ref{table4} and \ref{table5}. $A^{\rm nrel}_{1/2}$ according to Eq. (\ref{08}) 
involves the exact electron $g$-factor, and thus includes the leading QED
corrections. The relativistic corrections $A^{\rm rel}$ and $B^{\rm rel}$
have been calculated by Yerokhin in \cite{yer_hfs} in terms of $G_{M1}$ and $G_{E2}$
functions. $G_{M1}$ is defined by
\begin{equation}
A_J = \varepsilon\,\frac{Z^3}{8}\,\frac{m}{m_{\rm
    p}}\,\frac{\mu}{\mu_{\rm N}I}\,\frac{1}{3\,J\,(J+1)}\,G_{M1},
\end{equation}
where relativistic corrections to $G_{M1}$ are equal to $0.000015$ for $2^2P_{1/2}$,
$-0.000039$ for $2^2P_{3/2}$ states of Li, and 
$0.000153$ for $2^2P_{1/2}$, $-0.000161$ for $2^2P_{3/2}$ states of Be$^+$.
These number include also the so called negative-energy contributions.
$G_{E2}$ is related to $B_J$ coefficient by
\begin{equation}
B_{3/2} = \varepsilon\,m^2\,Q\,\frac{Z^3}{60}\,G_{E2},
\end{equation}
where relativistic corrections to $G_{E2}$ for $2^2P_{3/2}$ are equal to
$-0.000\,004$ in Li and $-0.000\,013$ in Be$^+$. These relativistic
corrections can in principle be evaluated within NRQED approach \cite{lihfs}
but so far we have not been able to obtain analytic formula for 
all Hylleraas integrals involved in matrix elements. 
The next to leading radiative (QED) correction $A^{\rm qed}_{1/2}$ 
(beyond the anomalous magnetic moment) is proportional to the Fermi 
contact interaction and is known from
hydrogenic atoms. In terms of the $H_a$ operator it is
\begin{equation}
H^{\rm qed}_a = H_a\,\frac{2}{g}\,Z\,\alpha^2\biggl(\ln 2-\frac{5}{2}\biggr).
\end{equation}
The last significant contribution is the finite nuclear size correction, 
the extended electric and magnetic distribution within nucleus.
It is given by the formula 
\begin{equation}
H^{\rm fns}_a = H_a\,(-2\,Z\,\alpha\,m\,r_Z),
\end{equation}
where
\begin{equation}
r_Z = \int d^3r\,d^3r' \rho_E(r)\,\rho_M(r')\,|\vec r -\vec r\,'|.
\end{equation}
Using exponential parametrization of electric and magnetic formfactors
\begin{eqnarray}
\rho_E(r) &=& \frac{3\,\sqrt{3}}{\pi\,r_E^3} e^{-2\,\sqrt{3}\,r/r_E},\\
\rho_M(r) &=& \frac{3\,\sqrt{3}}{\pi\,r_M^3} e^{-2\,\sqrt{3}\,r/r_M},
\end{eqnarray}
the Zemach radius $r_Z$ is
\begin{equation}
r_Z =\frac{35\,(r_E+r_M)^4+14\,(r_E^2-r_M^2)^2-(r_E-r_M)^4}{32\,\sqrt{3}\,(r_E+r_M)^3}.
\end{equation}
For all but $^{11}$Be nuclei we assume $r_E=r_M$, thus
\begin{equation}
r_Z =\frac{35\,r_E}{16\,\sqrt{3}} = 1.263\,r_E, 
\end{equation}
and take charge radii from the recent isotope shift measurements in Li \cite{lit_iso2}
and Be$^+$ \cite{be_iso} supplemented with isotope shift calculations in
\cite{lit_rel}.  For the Gaussian distribution one obtains \cite{yer_hfs}
$r_Z = 1.30\,r_E$ what demonstrates a weak dependence of $r_Z$
on an arbitrarily assumed shape of the charge distribution, with one exception.
The $^{11}$Be nucleus has a single neutron halo, what means
that $r_M$ is much larger than $r_E$ and the nuclear finite size becomes 
much larger. We employ here the result of direct calculations 
from \cite{fujita}, which is
\begin{equation}
H^{\rm fns}_a = H_a\,(-0.000\,717).
\end{equation}
At the same time the nuclear polarizability correction 
is also much larger and of the opposite sign to the finite size effect. 
Since it is very difficult to estimate, it will be neglected here.
The final results for $A_{1/2}$,$A_{3/2}$, and $B_{3/2}$ include the uncertainty
coming from the higher order corrections, which we estimate to be 
25\% of $A^{\rm qed}$ and the uncertainty due to the approximate
treatment of the nuclear structure which we estimate to be 25\%  of $A^{\rm fns}$
for all $A$ coefficients, while for the $B$ 
coefficients we assume the final uncertainty  to be  
the sum of 10\% of $B^{\rm rel}$ and 0.1\% of $\delta B$. \\

\section{Conclusions}
In comparison to experimental values we observe significant discrepancies
for the isotope shift in the fine structure, see Table \ref{table3}.
Although the theoretical fine structure of $^7$Li is consistent with
experimental values, the differences can be associated to $O(\alpha^2)$
relativistic corrections. The isotope shift, as it has already been noted in 
\cite{drake_fine,drake_be} differs significantly
between different experiments and theoretical predictions.
In view of the recent determination of the nuclear charge radii from 
the isotope shift of $2S_{1/2}-2P_{1/2}$ transition in Be$^+$ ions,
it is important to resolve these discrepancies. In this respect, 
we note the recent critical examinations \cite{noble2} of all experimental 
values of the fine structure and isotope shift measurements in $^6$Li and $^7$Li.
Considering hyperfine splittings we observe good agreement 
with the previous calculations of Yerokhin in \cite{yer_hfs}, 
particularly for the $A_{1/2}$ coefficients. Slight discrepancies
with experiments for the $A$ coefficients of the $2P$ state 
indicate that the magnetic moment obtained 
from the hfs measurement for the $2S$ state
may not be as accurate as claimed. This is because
the treatment of the nuclear structure corrections by the elastic charge and
magnetic formfactors is very approximate, and the accuracy of this 
approximation is not known. We think that the more accurate approach
shall  employ the effective nuclear Hamiltonian using the 
so called chiral perturbation theory. Then the nuclear structure
correction to the atomic hfs consists of the leading Low correction,
Zemach corrections from individual nucleons and the nuclear vector 
polarizability \cite{vecpol}. Unfortunately, the explicit calculations 
for nuclei with more than 3 nucleons is difficult and has not been performed so far.
Certainly the nuclear vector polarizability correction  
is significant for halo nuclei, and it would be worth to calculate it.
At present, without detailed knowledge of nuclear structure,
the determination of magnetic moments from atomic spectroscopy measurements
can be uncertain. Therefore, better accuracy can be achieved
when two measurements are combined in such a way, that this nuclear structure
correction, proportional to the Fermi interaction cancels out,
for example in $A_{1/2}+A_{3/2}$ of the $P$ state of Li  and Be$^+$.  
Theoretical accuracy for this combination is limited only by
higher order QED corrections and knowing both $A$ constants,
we shall be able to derive magnetic moments with relative precision of
about $10^{-5}$ without referencing to magnetic 
moments of stable isotopes, or with precision of 
the magnetic moment of the reference nucleus.

\section*{Acknowledgments}
Authors wish to acknowledge fruitful discussions
with Vladimir Yerokhin. This work was support by NIST
through Precision Measurement Grant PMG 60NANB7D6153.


\begin{thebibliography}{99}
\bibitem{7be_quad} The quadrupole moment of $^7$Be  has not been measured yet.
                   We use the theoretical result from \cite{arai}, as the
                   other theoretical calculations \cite{qube2} are not conclusive.
\end{thebibliography}
\end{document}